\documentclass[aps,twocolumn,groupedaddress,graphics]{revtex4}
\usepackage{graphics}
\usepackage{epsfig}
\begin{document}
\preprint{HEP/123-qed}
\bibliographystyle{apsrev}

\title{Electron Entanglement \em via \em  a Quantum Dot}

\author{W. D. Oliver}
\email[]{woliver@stanford.edu} \homepage[]{URL:
feynman.stanford.edu}
\author{F. Yamaguchi}
\author{Y. Yamamoto}
\thanks{also at NTT Basic Research Laboratories, 3-1 Morinosato-Wakamiya Atsugi, Kanagawa, 243-01 Japan}
\affiliation{Quantum Entanglement Project, ICORP, JST, \\
E. L. Ginzton Laboratory, Stanford University, Stanford, CA 94305}
\date{July 15, 2001}

\begin{abstract}
\vspace{0.2in}
This Letter presents a method of electron
entanglement generation. The system under consideration is a
single-level quantum dot with one input and two output leads. The
leads are arranged such that the dot is empty, single electron
tunneling is suppressed by energy conservation, and two-electron
virtual co-tunneling is allowed. This yields a pure, non-local
spin-singlet state at the output leads. Coulomb interaction is the
nonlinearity essential for entanglement generation, and, in its
absence, the singlet state vanishes. This type of electron
entanglement is a four-wave mixing process analogous to the photon
entanglement generated by a $\chi^{(3)}$ parametric amplifier.
\end{abstract}

\maketitle

Identical quantum particles are inherently indistinguishable, and
this oftentimes leads to non-classical behavior such as
entanglement in quantum mechanical systems. The
Einstein-Podolsky-Rosen (EPR) state
\cite{Einstein_etal35,Bohr38,Bohm54} is an interesting example of
two-particle entanglement, because it has potential use in secure
quantum communication protocol \cite{Bennett84,Ekert91}, quantum
information processing \cite{Bennett00}, and fundamental tests of
quantum mechanics \cite{Bell64}. Photons in nonlinear media
interact to produce polarization-entangled EPR pairs and have been
used in experimental demonstrations of quantum state teleportation
\cite{Bouwmeester_etal97,Boschi_etal98,Furusawa_etal98}, quantum
non-demolition measurements \cite{Grangier_etal98}, and violations
of Bell's inequality \cite{Aspect_etal81,Weihs_etal98}. Although
entanglement with ions \cite{Rowe01} and between atoms and cavity
field modes has been demonstrated
\cite{Hagley97,Maitre97,Nogues_etal99}, to our knowledge, there
have yet to be any experimental demonstrations specifically
utilizing EPR pair-type entangled electrons. Recently, there have
been several proposals to generate
\cite{Loss98,Burkard99,Barnes00,Recher01} and detect
\cite{Loss00,Burkard00,Oliver00,Maitre_etal00} entangled
electrons. Electrons have been demonstrated to have long spin
dephasing times in semiconductors \cite{Kikkawa97,Kikkawa99}. In
addition, the quantum optics tools
\cite{Buttiker92,Martin_Landauer92}, for example an electron
waveguide \cite{vanWees88,Wharam88}, beamsplitter
\cite{Liu_etal98,Oliver_etal99}, intensity interferometer
\cite{Oliver_etal99,Henny_etal99}, and collision analyzer
\cite{Liu_etal98,Oliver00}, required to detect entangled electrons
have been demonstrated in two-dimensional electron gas systems.
Furthermore, the lossless nature of electrons and the noiseless
property of a cryogenic Fermi source may provide experimental
advantages, for example high detection efficiency
\cite{Rowe01,Maitre_etal00}, over their photon counterparts.

\begin{figure}
\epsfig{figure=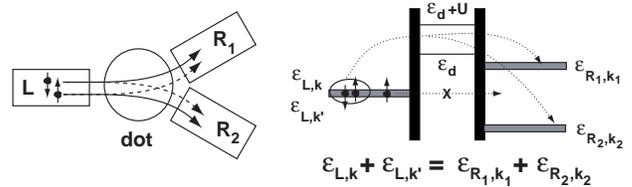,width=3.2in}
\caption{a) Three-port
quantum dot. b) Energy band diagram for three-port quantum dot
with non-degenerate energy leads which act as energy filters.
\label{Fig:fig1}}
\end{figure}

In this Letter, we consider a means to generate entangled EPR
pairs with electrons using a three-port quantum dot
(Fig.~\ref{Fig:fig1}a) operating in the coherent tunneling regime
\cite{Devoret91,Sukhorukov01}. The dot consists of a single input
lead and two output leads, with an energy band diagram shown in
Fig.~\ref{Fig:fig1}b. The lead arrangement is such that the dot is
empty. Conceptually, there are two key factors to the successful
operation of this entangler. The first is that the leads are
non-degenerate and of narrow width in energy, thus acting as
``energy filters''. Single electron tunneling does not conserve
energy and is forbidden. However, the lead energies can be
arranged such that two-electron co-tunneling events do conserve
energy, and thus the lowest order contribution to the tunneling
current is two-electron virtual co-tunneling through the dot. The
second is that double occupancy of the dot incurs a charging
energy. The Coulomb interaction mediates electron entanglement in
this system, as a nonlinear medium does for photon entanglement.
In its absence, the singlet state generation vanishes.
Fundamentally, this is because the system has a high degree of
symmetry. With the Coulomb interaction turned off, the singlet and
triplet states destructively interfere. However, in the presence
of Coulomb interaction, the system symmetry is reduced and the
singlet state destructive interference becomes imperfect, while
the triplet state destructive interference remains complete. This
leads to a net singlet state amplitude at the output of the dot.

Experimental verification of electron entanglement might be
achieved through an electron bunching/anti-bunching experiment
\cite{Burkard00,Oliver00}, or spin correlation measurements and,
ultimately, a Bell's inequality test \cite{Maitre_etal00}. Since,
to our knowledge, there is no demonstrated method to coincidence
count with electrons, an entangler ideally should have high
efficiency. That is, the output is a continuous and coherent
singlet state with no noise. Then, at least in principle, standard
noise measurement techniques could be adopted to infer the degree
and type of entangled state produced. Although outside the scope
of this Letter, we consider elsewhere the issue of resonance
conditions for the quantum dot entangler to enhance detection
\cite{Oliver01}.


The quantum dot is analyzed in the coherent tunneling regime using
the Anderson Hamiltonian with an on-site Coulomb energy term U. We
consider only a single, spin-degenerate energy level for the dot,
and there are no single electron excitations within the dot.
\begin{eqnarray}
  \label{Anderson}
  \hat{H}_{\textrm{And}} & = &
  \sum_{\eta,k,\sigma}\varepsilon_{\eta,k}\hat{a}_{\eta,k,\sigma}^{\dag}\hat{a}_{\eta,k,\sigma}
  +
  \sum_{\sigma}\varepsilon_{d}\hat{c}_{\sigma}^{\dag}\hat{c}_{\sigma} \nonumber \\
  & & + U\hat{n}_{\uparrow}\hat{n}_{\downarrow}
   + \sum_{\eta,k,\sigma}
  \left(V_{\eta}\hat{a}_{\eta,k,\sigma}^{\dag}\hat{c}_{\sigma} + h.c. \right)
\end{eqnarray}
where $\eta\in\{L,R_{1},R_{2}\}$ is the lead label, $k$ is the
lead electron momentum, $\sigma \in\{\uparrow,\downarrow\}$ is the
electron spin, $V_{\eta}$ is the overlap matrix element between
the dot and the lead states, $\hat{a}$ ( $\hat{a}^{\dag}$) is the
annihilation (creation) operator for the lead electrons, $\hat{c}$
($\hat{c}^{\dag}$) is the annihilation (creation) operator for the
dot electrons, and $\hat{n}_\sigma \equiv
\hat{c}_\sigma^{\dag}\hat{c}_\sigma$ is the dot electron number
operator. The dot energy levels, $\varepsilon_d$ and
$\varepsilon_d+U$, in Fig.~\ref{Fig:fig1} are taken to be off
resonance with the leads. The left lead energy is below its
quasi-Fermi level so that the lead is full of electrons. The right
leads are empty. In addition, we set $\varepsilon_d = 0$, thereby
referencing all energies to $\varepsilon_d$. The lead energies
$\varepsilon_{\eta,k}$ and the charging energy $U$ are left as
parameters which can be adjusted to consider different dot
configurations.

In all cases, the three-port quantum dot is biased such that
single electron tunneling from the left lead to the right lead is
suppressed, that is, $\varepsilon_{L,k} \neq
\varepsilon_{R_1,k_1}\neq \varepsilon_{R_2,k_2}$. The lead energy
level widths are narrow enough that energy overlap between
different leads does not occur. However, two-electron virtual
co-tunneling does conserve energy, that is, $\varepsilon_{L,k} +
\varepsilon_{L,k'}= \varepsilon_{R_1,k_1} +
\varepsilon_{R_2,k_2}$, requiring one electron from the left lead
to go to lead $R_1$ and the other to go to lead $R_2$. This is the
energy conserving process considered throughout this Letter.

In terms of a perturbation expansion in the tunneling matrix
element $V$, the lowest order contribution to the current from the
left lead to the right leads is $\mathcal{O}\left(V^4\right)$.
Given the assumption that only one dot energy level is relevant in
this quantum dot, all higher-order terms contribute to either the
self-energy of the electrons or are higher orders of two-electron
co-tunneling. Therefore, a two-electron initial state is used in
this model,
\begin{equation}
  \label{initialstate}
  |\phi_i\rangle=\hat{a}_{L,k,\sigma}^{\dag}\hat{a}_{L,k',\sigma'}^{\dag}|0\rangle
\end{equation}
where $|0\rangle$ is the zero-particle state of this model system.
This initial state is an arbitrary selection of two electrons from
the entire left lead $T \approx 0$ ground state. Physical
quantities, such as current, can be found by summing over all
possible two-electron input and output states.

We introduce the notation $E_L$, $\Delta_L$, and $\Delta_R$ to
parameterize $\varepsilon_{L,k}$, $\varepsilon_{L,k'}$,
$\varepsilon_{R_1}$, and $\varepsilon_{R_1}$ and simplify the
presentation.
\begin{eqnarray}
E_L & \equiv &  \frac{1}{2}
                \left(\varepsilon_{L,k} + \varepsilon_{L,k'}\right)
               = \frac{1}{2}
                \left(\varepsilon_{R_1,k_1} +
                \varepsilon_{R_2,k_2}\right) \\
\Delta_L & \equiv &  \frac{1}{2}
                \left(\varepsilon_{L,k} - \varepsilon_{L,k'}\right)
                 \\
\Delta_R & \equiv &  \frac{1}{2}
                 \left(\varepsilon_{R_1,k_1} -
                \varepsilon_{R_2,k_2}\right)
\end{eqnarray}
The two-electron initial state has energies
$\varepsilon_{L,k},\varepsilon_{L,k'} = E_L \pm \Delta_L$ within
the left lead, and the two-electron final state has energies
$\varepsilon_{R_1,k_1},\varepsilon_{R_2,k_2} = E_L \pm \Delta_R$
in the right leads. The energy widths of the leads are denoted as
$\delta_L$ for the left lead and $\delta_R$ for each of the two
right leads. Suppressing single electron tunneling requires that
$\Delta_L < \Delta_R$ for any $\Delta_L$ and $\Delta_R$. In
practice, this would mean that $\delta_L$, $\delta_R << \Delta_R$.

\begin{figure}
\begin{center}
\epsfig{figure=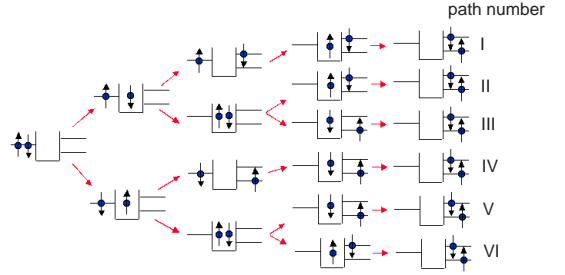,width=2.8in} \caption{Multiple
paths by which two electrons can virtually co-tunnel from the left
lead through an empty dot to the two right leads.
\label{Fig:fig3}}
\end{center}
\end{figure}

We consider first the case of a spin-up and spin-down electron
tunneling through the dot to the output leads as indicated in the
Fig.~\ref{Fig:fig1}. The initial state is Eqn.~\ref{initialstate}
with $\sigma = \uparrow$ and $\sigma' = \downarrow$. The
time-ordering operator in the perturbative expansion
\cite{Rodberg67,Mahan90} leads to six paths comprising twelve
unique time-orderings by which these two electrons can virtually
co-tunnel through the quantum dot. Each path has two
time-orderings, one as shown in Fig.~\ref{Fig:fig3} and another
due to the exchange of the output leads $R_1$ and $R_2$. This
interchange of $R_1$ and $R_2$ introduces a minus sign due to the
electron commutation relation when the state is written in its
``normal ordering'' form. Due to the coherent nature of the
virtual tunneling process, the six paths will interfere to produce
the resulting final state output.
The six paths are presented in terms of a singlet and a triplet
contribution using the shorthand notation $|S\rangle , |T\rangle
\equiv  \left(
     \hat{a}_{R_1\uparrow}^{\dag}\hat{a}_{R_2\downarrow}^{\dag}
     \mp \hat{a}_{R_1\downarrow}^{\dag}\hat{a}_{R_2\uparrow}^{\dag}
     \right)|0\rangle$
to indicate the singlet and triplet states. The factor $C \equiv
V_{L}^{*2}V_{R_1}V_{R_2} \frac{\exp \left[-\frac{i}{\hbar}
     \left(2E_L-\varepsilon_{R_1,k_1}-\varepsilon_{R_2,k_2} \right)
     \right]}{\left( 2E_L-\varepsilon_{R_1,k_1}-\varepsilon_{R_2,k_2}\right)}$
is common to all terms and
omitted until the end.
\begin{eqnarray}
     |\phi_I\rangle & = &
            \frac{ (\Delta_R^2-\Delta_L E_L)|S\rangle
            -\Delta_R (E_L-\Delta_L) |T\rangle }
                {(E_L-\Delta_L)(E_L^2-\Delta_R^2)(\Delta_L^2-\Delta_R^2)} \\
     |\phi_{II}\rangle & = &
            \frac{E_L |S\rangle -\Delta_R(E_L-\Delta_L) |T\rangle }
                {(E_L-\Delta_L)(E_L^2-\Delta_R^2)(2E_L-U)} \\
    |\phi_{III}\rangle & = &
            \frac{E_L |S\rangle + \Delta_R (E_L-\Delta_L) |T\rangle }
                {(E_L-\Delta_L)(E_L^2-\Delta_R^2)(2E_L-U)} \\
    |\phi_{IV}\rangle  & = &
            \frac{(\Delta_R^2 + \Delta_L E_L) |S\rangle
            + \Delta_R (E_L+\Delta_L) |T\rangle }
                {(E_L+\Delta_L)(E_L^2-\Delta_R^2)(\Delta_L^2-\Delta_R^2)} \\
    |\phi_V\rangle& = &
            \frac{E_L |S\rangle +\Delta_R (E_L+\Delta_L) |T\rangle   }
                {(E_L+\Delta_L)(E_L^2-\Delta_R^2)(2E_L-U)} \\
     |\phi_{VI}\rangle & = &
            \frac{E_L |S\rangle - \Delta_R (E_L+\Delta_L) |T\rangle }
                {(E_L+\Delta_L)(E_L^2-\Delta_R^2)(2E_L-U)}
\end{eqnarray}
The output states from paths $I$ and $IV$ do not incur a charging
energy $U$, while the remaining states do incur a charging energy
$U$ during the virtual tunneling event.

It is now clear how the triplet states destructively interfere.
Paths $I$ and $IV$, paths $II$ and $III$, and paths $V$ and $VI$
have triplet contributions which cancel. Only singlet states
remain, and the singlet contributions along paths $II$ and $III$
and along paths $V$ and $VI$ are identical. Furthermore, paths
$I$, $II$, and $III$ contain the factor $(E_L-\Delta_L)$ in the
denominator, indicating that the first electron to enter the dot
along these paths was the $L,k$ electron. Conversely, paths $IV$,
$V$, and $VI$ contain the factor $(E_L+\Delta_L)$ in the
denominator, indicating that the first electron to enter was the
$L,k'$ electron. Combining paths according to the electron which
initially enters the dot yields the following expressions. The
triplet state contributions of paths $I$ and $IV$ are simply
written as $\pm \tau(E_L, \Delta_L, \Delta_R)$, since they cancel
upon summation as stated above.

\begin{widetext}
\begin{eqnarray}
 |\phi_{I}\rangle + |\phi_{II}\rangle  + |\phi_{III} \rangle  & = &
            \frac{- U \left( \Delta_R^2 - \Delta_L E_L \right) - 2 \Delta_L E_L \left( E_L - \Delta_L \right)}
                {(E_L-\Delta_L)(\Delta_L^2-\Delta_R^2)(E_L^2-\Delta_R^2)(2E_L-U)}|S\rangle
                + \tau(E_L, \Delta_L, \Delta_R) |T\rangle \\
|\phi_{IV}\rangle + |\phi_{V}\rangle  + |\phi_{VI} \rangle & = &
            \frac{- U \left( \Delta_R^2 + \Delta_L E_L  \right) + 2 \Delta_L E_L \left( E_L + \Delta_L \right)}
                {(E_L+\Delta_L)(\Delta_L^2-\Delta_R^2)(E_L^2-\Delta_R^2)(2E_L-U)}|S\rangle
                - \tau(E_L, \Delta_L, \Delta_R) |T\rangle
 \end{eqnarray}
 \end{widetext}
The interference of the singlet states is now clear. In the
absence of the charging energy $U$, the singlet states in paths
$I$, $II$, and $III$ destructively and completely interfere with
those in paths $IV$, $V$, and $VI$. However, the presence of the
charging energy $U$ adds an additional contribution to each set of
paths. This additional $U$-dependent contribution contains a part
which still destructively interferes between the two sets of
paths, but it also contains a part which constructively
interferes, leaving a residual singlet state at the output.
Reinserting the common factor $C$, the final result for the total
output state is
\begin{eqnarray}
\label{nzeroresult2}
     |\psi(t) \rangle  =
     \frac{2CE_L}{\left(E_L^2-\Delta_R^2\right) \left(E_L^2-\Delta_L^2 \right)}
     \frac{U}{2E_L-U}|S\rangle.
 \end{eqnarray}

The remaining case is the virtual co-tunneling of two electrons
with the same spin. The input state is Eqn.~\ref{initialstate}
with $\sigma = \sigma'$. Due to the Pauli exclusion principle,
only paths $I$ and $IV$ can contribute to the output state
amplitude. Since there are no singlet states in the same spin case
and the triplet states destructively interfere for paths $I$ and
$IV$, there is no same-spin co-tunneling contribution at the
output.

The conclusion is that virtual co-tunneling through an empty
quantum dot in the presence of Coulomb interaction generates
entangled electron spin-singlet states. The physical meaning of
equation~\ref{nzeroresult2} can be explained in the following way.
Two electrons (with possibly different energies) virtually tunnel
through the quantum dot along six different paths (twelve unique
time-orderings). These different paths interfere with each other.
Two of the paths ($I$ and $IV$) do not include double occupancy of
the dot and, thus, incur no charging energy. The remaining four
paths do incur a charging energy due to a virtual double occupancy
of the dot. The interference for the triplet states is destructive
and complete. That is, the triplet state contribution of each path
has an opposite-parity partner with which it destructively
interferes. The result is independent of the charging energy $U$,
since the triplet states of those paths which do depend on
charging energy destructively self-interfere, and those paths
which do not self-interfere also. There is no requirement to
``mix'' the $U$-dependent and $U$-independent paths to get the
complete destructive interference of the triplet states.

However, the system symmetry or ``non-mixing property'' is broken
in the case of the singlet states. The interference is destructive
for the singlet states, but only complete when the charging energy
$U=0$. For $U\neq0$, the destructive interference becomes
imperfect, and residual singlet state remains at the output. One
perspective is that the interference occurs between paths which
share the same electron in the first tunneling event. For example,
there are three paths which share the $L,k$ electron first
tunneling into the dot. Call the $L,k$ electron paths set $A$ and
the remaining three $L,k'$ electron paths set $B$. Within each
set, two of the three paths depend on $U$ and one does not.
Therefore, within each set, there is a mixing of paths which do
depend and do not depend on $U$. In the limit $U=0$, the
contributions of set A and set B are equal in magnitude and of
opposite parity; they cancel when summed. However, $U\neq0$ breaks
this symmetry, creating a residual $U$-dependent contribution that
does not cancel when sets A and B are summed. Admittedly, the
choice of grouping paths into sets A and B, while physically
compelling, is nonetheless arbitrary. It may be more appropriate
to state that $U$ breaks the system symmetry, leading to a net
singlet contribution.

The physical meaning of the same-spin case is similar. The
same-spin electrons can only form a spin-triplet state at the
output and, therefore, destructively interfere completely.
Furthermore, the same-spin case only traverses the dot along paths
$I$ and $IV$, which do not incur any charging energy. In this
sense, there is no non-linear interaction to create an entangled
state at the output, and so, from this perspective also, the
destructive interference is complete.

The nonlinearity in this system, the Coulomb interaction as
manifest in the charging energy $U$, is necessary to observe
entanglement at the output. The optical analogue to this system is
a $\chi^{(3)}$ parametric amplifier with a vacuum state as its
input. Two input photons, created through spontaneous emission,
interact within the $\chi^{(3)}$ nonlinear medium and generate an
entangled pair of output photons. This is a four-wave mixing
process; two photons in and two photons out. The $\chi^{(3)}$
nonlinearity mediates the entanglement in the photon case, whereas
it is the Coulomb interaction in the electron case. This
correspondence between nonlinear optics and Coulomb-mediated
wave-mixing may be applicable to other quantum dot configurations
and other electron or composite particle systems that exhibit
Coulomb charging behavior.

The authors gratefully acknowledge support from MURI and JSEP. W.
D. O. gratefully acknowledges additional support from MURI and the
NDSEG Fellowship Program.

\end{document}